\documentclass[iop]{emulateapj}
\usepackage{float}
\usepackage{amsmath}
\usepackage{booktabs}
\usepackage{booktabs,threeparttable}
\usepackage{makecell}
\usepackage{threeparttable, tablefootnote}
\usepackage{changes}
\usepackage{lipsum}
\definechangesauthor[name={Per cusse}, color=orange]{per}
\setremarkmarkup{(#2)}

\slugcomment{Submitted to \apj, November 2017}

\shorttitle{Observations of the Abell~3395/Abell~3391 Intercluster Filament}
\shortauthors{G. E. Alvarez, S. W. Randall, H. Bourdin, C. Jones, \& K.H. Bockelmann}

\begin{document}
\nocite{*}

\title{{\it Chandra} and {\it XMM-Newton} Observations of the Abell~3395/Abell~3391 Intercluster Filament}

\author{Gabriella E. Alvarez \altaffilmark{1}, 
Scott W. Randall, Herv{\'e} Bourdin, Christine Jones, and Kelly Holley-Bockelmann\altaffilmark{1}\altaffilmark{2}}
\affil{{Harvard-Smithsonian Center for Astrophysics,
    60 Garden Street, Cambridge, MA 02138}}
\altaffiltext{1}{{Astrophysics Department, Vanderbilt University, Nashville, TN 37235}}
\altaffiltext{2}{{Fisk University, Nashville, TN, 37208}}
\email{gabriella.alvarez@cfa.harvard.edu}

\begin{abstract}
We present \textit{Chandra} and \textit{XMM-Newton} X-ray observations
of the Abell 3391/Abell 3395 intercluster filament. It has been suggested that the galaxy clusters Abell 3395, Abell 3391, and the galaxy group ESO-161 located between the two clusters, are in alignment along a large-scale intercluster filament. We find that the filament is aligned close to the plane of the sky, in contrast to previous results. We find a global projected filament temperature kT = $4.45_{-0.55}^{+0.89}$~keV, electron density $n_e=1.08^{+0.06}_{-0.05} \times 10^{-4}$~cm$^{-3}$, and $M_{\rm gas} = 2.7^{+0.2}_{-0.1} \times 10^{13}$~M$_\odot$. The thermodynamic properties of the filament are consistent with that of intracluster medium (ICM) of Abell 3395 and Abell 3391, suggesting that the filament emission is dominated by ICM gas that has been tidally disrupted during an early stage merger between these two clusters. We present temperature, density, entropy, and abundance profiles across the filament. We find that the galaxy group ESO-161 may be undergoing ram pressure stripping in the low density environment at or near the virial radius of both clusters due to its rapid motion through the filament.
\end{abstract}


\section{Introduction} \label{sec:intro}

Theory and observations both suggest that there are fewer baryons detected in the local universe than predicted. Observations of the cosmic microwave background \citep[i.e.][]{planck2015}
 and Big Bang nucleosynthesis (BBN) models \citep{kaplinghat} predict that baryons comprise approximately $5\%$ of the total mass budget in the Universe. In the local universe, the known baryon content falls short by about a factor of two \citep{fuku,cenost,bregman,sinha}. This
 discrepancy is known as the ``missing baryons problem.'' 
It is theorized \citep[e.g.][]{cenost,dave}) that the bulk of these missing baryons may be in the form of a diffuse gas that traces the filaments of the cosmic web, known as the warm-hot intergalactic medium (WHIM). Simulations predict WHIM temperatures of 
$\log{T} \simeq 5 - 7$~K
and baryonic densities of  $n_b \simeq 10^{-6} - 10^{-5}$~cm$^{-3}$ \citep[see][]{bregman}, making this medium very difficult to observe directly with
existing observatories.
Evidence has been found for the WHIM in the form of absorption lines
in the soft X-ray spectra of high redshift objects
\citep[e.g.][]{zappacosta}. However, observations via absorption lines 
have been unable to constrain the amount of baryons present or to
trace large-scale filamentary structure. There has yet to be a high significance
observation of the WHIM in large-scale filaments 
\citep[i.e.][]{kull,fang} aside from a handful of possible detections of the more dense part of
the WHIM at the outskirts of galaxy clusters \citep[][]{wang,werner,eckert2015,bulbul}. More recently, \cite{graaff} claimed a $5.1\sigma$ detection of WHIM filaments using stacked Sunyaev Zel'dovich measurements.

Galaxy clusters are excellent probes of the large-scale distribution of the WHIM, because they are found at the intersection of dark matter filaments 
\citep[i.e.][]{gonzalez}. This
makes galaxy clusters excellent probes to study large-scale and intercluster filaments. The WHIM may have an impact on the intracluster medium (ICM) of galaxy clusters, particularly where the ICM in the outskirts of the clusters are expected to interface with the WHIM in large-scale filaments. Entropy profiles of the ICM are generally found to lie below what one would expect based on pure gravitational collapse models \citep{kaiser,voit} near a cluster's virial radius \citep[i.e.][]{edge,david,allen,arnaud,walker2013,urban}. This interaction triggers thermodynamic processes, causing departures from the expected hydrostatic equilibrium \citep[i.e.][]{ichikawa}.
One such process that could explain the observed entropy flattening is unresolved cool clumps of infalling gas at large
cluster radii \citep[][]{simionescu,tchernin}, but the required clumping factors in observations are
often larger than what is predicted by simulations
\citep[][]{walker2013,urban}. Other proposed mechanisms for observed entropy flattening include: accretion shocks that weaken as the cluster grows
\citep[][]{lapi,fusco}, non-thermal pressure support from
bulk motions, turbulence or cosmic-rays \citep[][]{lau,vazza,battaglia}, and electron-ion non-equilibrium \citep[][]{fox,wong2009,hoshino,avestruz}. All of these mechanisms are expected to correlate with large-scale structure filaments.  Thus, entropy flattening may indicate regions where the outskirts of clusters interface with WHIM filaments.

\begin{table*}[t]
\centering
\caption{Physical properties for the objects studied in this work. }
\label{tab:objectinfo}
\begin{tabular}{@{}ccccccc@{}}
\toprule
Object  & RA                                    & DEC                                  & $z$    & $T_X$ {[}keV{]} & $r_{500}$ {[}Mpc{]} & $M_{500}$ {[}M$_\odot${]}             \\ \midrule
A3391   & $06^{\rm{h}}26^{\rm{m}}22^{\rm{s}}.8$ & $-53^{\rm{d}}41^{\rm{m}}44^{\rm{s}}$ & 0.0551 & 5.39            & 0.90                & $2.16 \times 10^{14}$ \\
A3395   & $06^{\rm{h}}27^{\rm{m}}14^{\rm{s}}.4$ & $-54^{\rm{d}}28^{\rm{m}}12^{\rm{s}}$ & 0.0506 & 5.10            & 0.93                & $2.4 \times 10^{14}$  \\
ESO-161 & $06^{\rm{h}}26^{\rm{m}}05^{\rm{s}}.2$  & $-54^{\rm{d}}02^{\rm{m}}04^{\rm{s}}$ & 0.0520 & 1.09            &  0.51                & $2.8 \times 10^{13}$  \\ \bottomrule
\end{tabular}
\end{table*}

It is worth mentioning that there are exceptions to this entropy flattening
trend. For example, \cite{bulbul} found that the entropy profile of
Abell 1750 is consistent with a self-similar appearance near the virial radius, and argue that lower mass systems are less likely to exhibit entropy flattening. Subsequently, the same suggestion was made independently by \cite{tholken} that low mass systems are less likely to show evidence for flattened entropy profiles. In addition, this view is supported by the observations of the low mass fossil cluster RX J1159+5531 \citep{su}, which appears to adhere to self-similarity azimuthally. More relaxed clusters seem to follow self-similarity more closely than clusters undergoing mergers \citep[][]{eckert2012,eckert2013} (for a review see \cite{wong2016}).

The double peaked cluster Abell 3395 (hereafter A3395) was first characterized
with Einstein observations \citep{forman}. A3395 is relatively close, both in projected separation on the sky and in redshift, to
Abell 3391 (hereafter A3391). There is also a galaxy group ESO 161-IG 006 (hereafter ESO-161) located between the two subclusters in the intercluster filament, in alignment with the clusters. In Table~\ref{tab:objectinfo}, the cluster masses \citep{piffaretti} and group mass estimated in this work, redshifts for the group and clusters \citep{tritton,santos}, positions\footnote[1]{The NASA/IPAC Extragalactic Database (NED)
is operated by the Jet Propulsion Laboratory, California Institute of Technology,
under contract with the National Aeronautics and Space Administration.}, the radius at which the mean cluster density is 500 times the critical density of the universe at the redshift of the clusters \citep{piffaretti}, and their X-ray temperatures \citep{vikhlinin} are shown. The cluster centers have a separation of 
$47.08\arcmin$ 
on the sky, which corresponds to 2.9 Mpc at their mean redshift. {\it ASCA}, {\it ROSAT}, {\it Planck}, and {\it Suzaku} observations indicate and confirm that A3395 and A3391
are connected by an intercluster filament, with detectable diffuse
emission apart from cluster scattered light \citep{tittley,planck2013,sugawara}. Previous dynamical analysis suggests that the filament is aligned almost parallel to the line of sight, with an inclination angle in the $3.1^\circ-9^\circ$
range \citep{tittley}. Thus, the filament is an ideal target for direct detection of
the diffuse gas since the projected surface brightness is much higher than if the system were perpendicular to the line of sight. 

Here, we report findings based on six observations
with {\it Chandra} and {\it XMM-Newton} of A3395, A3391, and the
connecting filament.
This paper is organized as follows: in Section 2 we discuss the data
reduction and analysis techniques; in
Section 3 we present the resulting images, spectra, temperature, and metallicity profiles; in Section 4 we
discuss the nature and orientation of the filament as well as ESO-161; and conclusions and a summary of this work are presented in Section 5. Unless otherwise stated, all uncertainties are 90\% confidence
intervals. For this analysis we assume the abundance table of
\cite{grevesse}. The mean redshift of A3395 and A3391 is $\bar{z}=0.053$, 
such that 1\arcsec\ on the sky corresponds to $\approx 1.04$ kpc. We use the fiducial cosmology $H_0 = 70 \rm{\ km \ s^{-1} \  Mpc^{-1}}$, $\Omega_M = 0.3$, and $\Omega _\Lambda = 0.7$.

\section{Data Analysis} \label{sec:da}

This section discusses the data reduction and analysis techniques employed in this work for \textit{Chandra} and \textit{XMM-Newton}.

\subsection{Chandra Data Reduction and Analysis} \label{subsec:chdra}

\begin{table*}[t]
\centering
\caption{Summary of the {\it Chandra} and {\it XMM-Newton} X-ray pointings}
\label{tab:pointings}
\begin{tabular}{@{}cccccccc@{}}
\toprule
Observatory  & Pointing        & ObsID      & RA        & Dec          & Date Obs   & \begin{tabular}[c]{@{}c@{}}Exposure {[}ks{]}\\ ACIS-I \\EMOS1/EMOS2/EPN\end{tabular} & PI            \\ \midrule
{\it Chandra}    & A3391           & 4943       & $06^{\rm{h}}26^{\rm{m}}22^{\rm{s}}.20$ & $-53^{\rm{d}}41^{\rm{m}}37^{\rm{s}}.50$ & 2004-01-15 & 18.3                                                                            & T. Reiprich   \\
\textit{Chandra}    & Filament North  & 13525      & $06^{\rm{h}}25^{\rm{m}}22^{\rm{s}}.52$ & $-53^{\rm{d}}53^{\rm{m}}54^{\rm{s}}.09$ & 2012-08-18 & 48.4                                                                            & S. Randall    \\
\textit{Chandra}    & Filament Center & 13519      & $06^{\rm{h}}26^{\rm{m}}10^{\rm{s}}.69$ & $-54^{\rm{d}}05^{\rm{m}}08^{\rm{s}}.53$ & 2012-08-17 & 47.1                                                                            & S. Randall    \\
\textit{Chandra}    & Filament South  & 13522      & $06^{\rm{h}}26^{\rm{m}}46^{\rm{s}}.24$ & $-54^{\rm{d}}17^{\rm{m}}05^{\rm{s}}.87$ & 2012-08-12 & 48.8                                                                            & S. Randall    \\
\textit{Chandra}    & A3395           & 4944       & $06^{\rm{h}}26^{\rm{m}}49^{\rm{s}}.56$ & $-54^{\rm{d}}32^{\rm{m}}35^{\rm{s}}.16$ & 2004-07-11 & 20.7                                                                            & T. Reiprich   \\
\textit{XMM-Newton} & Filament Center & 0400010201 & $06^{\rm{h}}26^{\rm{m}}31^{\rm{s}}.62$ & $-54^{\rm{d}}04^{\rm{m}}44^{\rm{s}}.7$  & 2007-04-06 &        38.2/38.5/23.1                                                                         & M. Henriksen \\ \bottomrule
\end{tabular}
\end{table*}

\begin{figure*}[ht!]
    \centering
    \includegraphics[width=\textwidth]{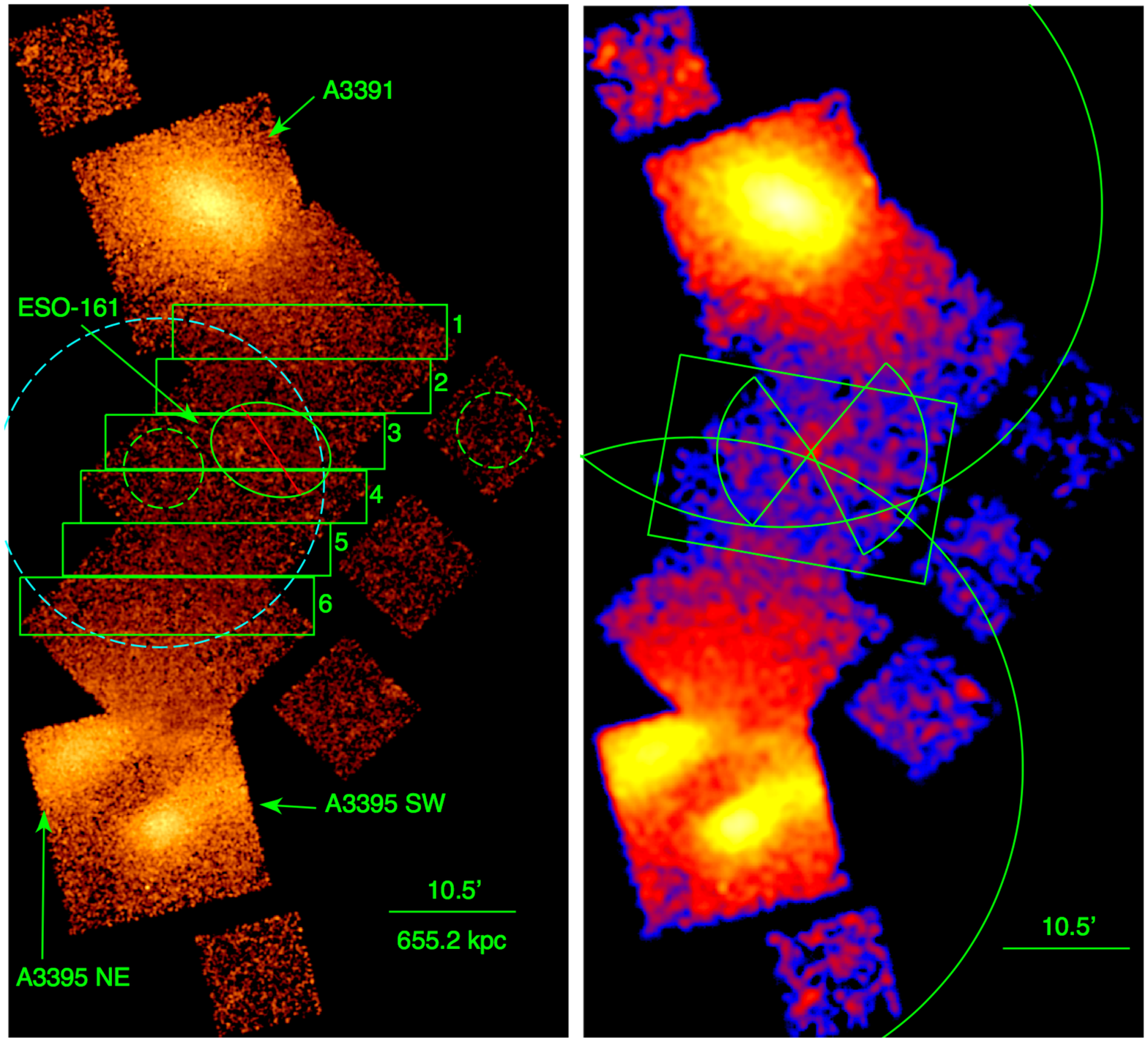}
    \caption{Left: The annotated background-subtracted exposure-corrected mosaic {\it Chandra} image of A3391, A3395, and the intercluster filament is shown in the 0.3-7.0 keV energy band and smoothed by a 12\arcsec~gaussian. The boxes denote regions used for the temperature profile of the filament. The excluded region, marked by the ellipse and red line, contains the galaxy group ESO-161. The green dashed circular region to the east of ESO-161 is used for local background modeling for the group temperature measurement. The green dashed circular region to the west on ACIS-I6 is used for background in all {\it Chandra} spectral analyses. The {\it XMM-Newton} (see Table~\ref{tab:pointings}) field of view is shown in the dashed cyan region for reference. Right: Same as left but smoothed  by a 40\arcsec gaussian to highlight the intercluster filament. Spectra extracted from the green box region were used to estimate the global temperature and density of the filament. The northern and southern circles are $r_{200}$ for A3391 and A3395 respectively. The wedges are used to derive the surface brightness profile of the group.}
    \label{fig:mosaic}
\end{figure*}

A summary of the observations is given in
Table~\ref{tab:pointings}. The aimpoint for each Chandra observation
was on the front-side illuminated ACIS-I CCD. We use CIAO version 4.8
and CALDB 4.7.2 to reduce the data to level 2 event files 
with the {\em chandra\_repro} script. The observations were
taken in very faint (VF) mode and the  
event
and background files were filtered appropriately.  We use 
the CIAO tool {\em deflare} to remove periods of strong flaring or data drop outs by removing
periods where the light curve is more than $3\sigma$ from the mean.
We find no instances of strong flaring. The total filtered ACIS-I
exposure time is 183.3 ks. We chose blank sky background files closest to the period of observation for each {\it Chandra} observation listed in Table~\ref{tab:pointings} for imaging. We use the CIAO tool {\em reproject\_events} to create images
with the blank sky background files for all of the observations. 
These background images were normalized to match the hard band
(10-12~keV) count rate in the observations to account for differences
in the particle background.
We create exposure maps for each image by utilizing the {\em asphist} and {\em mkinstmap} routines. We used these to create a background
subtracted, exposure corrected mosaic image in the 0.3-7.0 keV band, shown in Figure~\ref{fig:mosaic}.

To find background point and extended sources, we use the CIAO tool
{\em wavdetect} with 
wavelet 
scales 
of 2, 4, 6, 8, 12, and 16 pixels, where the pixels are 
0.98\arcsec\
in length.
These sources are then excluded from all spectra, surface brightness profiles, and images.
For the purposes of making images, we use the CIAO tool {\em dmfilth} to
fill in regions of excluded point sources in all of the observations by drawing photons from a
Poisson distribution matched to local background regions around the point sources. 

The {\em specextract} procedure with CIAO was used to extract spectra as well as the appropriate response files for analysis. All spectra in this work were grouped to a minimum of 40 net counts per bin.

XSPEC version 12.9.0 was used to perform the spectral analysis. Rather than use the CALDB blank sky files for background modeling, we use the circular westernmost
region on the ACIS-I6 chip shown in the left panel of Figure~\ref{fig:mosaic}. This has advantages over the blank sky background files, which are a sky average rather than the background in a nearby region of the sky. The stowed {\it Chandra}
background files, in which long exposures were taken with ACIS stowed and in VF mode, are used for instrumental background in the spectral fitting with an applied hard-band (10-12 keV) correction as was done for the blank sky background files for imaging. The scaled stowed spectra are consequently subtracted from the source spectra and local I6 background spectrum during spectral modeling. The stowed dataset accurately represents the quiescent, non-X-ray background (NXB), and this dataset introduces an additional $\pm 2\%$ statistical uncertainty (for more information on the stowed dataset we refer the reader to \cite{hickox}). For our faintest region (see Section~\ref{subsec:spectroscopy}), the effect that the systematic NXB uncertainty has on our measured kT and XSPEC normalization error range is an increase of less than $10\%$ and less than $3\%$ respectively. This impact on the error ranges is small for our faintest region, so we do not include the systematic NXB uncertainty in our error budget. The background spectrum on the I6 chip (see left panel of Figure~\ref{fig:mosaic}) is simultaneously fit with the source spectra to include background uncertainties in the calculated error ranges. A \textit{RASS} spectrum was also extracted from an annulus with an inner radius of $1^{\circ}$ and an outer radius of $1.1^{\circ}$ centered around ESO-161 from the \textit{RASS} observation of this system in order to better constrain the local background parameters in our spectral fits. This spectrum was simultaneously fit as part of the background model throughout the {\it Chandra} analysis. An absorbed Astrophysical Plasma Emission Code (APEC) \citep{smith} model was used for the source spectra.  The background spectra were simultaneously fit, along with the source spectra with an absorbed powerlaw (PL) for the cosmic X-ray background, an absorbed APEC model for the galactic halo (GH), as well an unabsorbed APEC model for the local hot bubble (LHB). The ICM, GH, and CXB are absorbed assuming a Galactic Hydrogen density column of $N_H=6.3 \times 10^{20}$ cm$^{-2}$ found with the ftool {\em nh} \citep{kalberla}. The LHB is unabsorbed. Parameters from the background spectral fit are shown in Table~\ref{tab:bg}.

\begin{table}[ht!]
\centering
\begin{threeparttable}[b]
\caption{CXB and foreground components derived from a spectral fit of RASS data as well as Chandra data from the green dashed region on the ACIS-I6 chip (see the left panel of Figure~\ref{fig:mosaic})without contamination from source data.}
\label{tab:bg}
\begin{tabular}{@{}lll@{}}
\toprule
Component & \begin{tabular}[c]{@{}l@{}}kT/$\Gamma$ \\ {[}keV{]}/$\Gamma$ \end{tabular} & \begin{tabular}[c]{@{}l@{}}$N$ {[}cm$^{-5}$/photons \\ keV$^{-1}$~cm$^{-2}$~s$^{-1}$ at 1 keV{]}\\ (\textit{Chandra/RASS})\end{tabular} \\ \midrule
LHB       & $0.092^{+0.06}_{-0.05}$                                                     & \makecell{$1.55^{+8.0}_{-0.17}\times10^{-5}$}                                                                    \\
GH        & $0.37^{+0.31}_{-0.09}$                                                     & \makecell{$2.2^{+0.76}_{-0.92}\times10^{-5}$}                                                                    \\
PL        & $1.4^f$                                                                       & \makecell{$2.7^{+0.55}_{-0.58}\times10^{-5}$/$2.1^{+0.28}_{-0.35}\times10^{-3}$}                                               \\ \bottomrule
$f$ fixed parameter \\
$\Gamma$ photon index
\end{tabular}
\end{threeparttable}
\end{table}
All spectra are fit in the energy range 0.5-7.0 keV for \textit{Chandra} data, 0.3-12.0 keV for \textit{XMM} data, and 0.1-2.0 keV for the \textit{RASS} data. All fitted parameters (temperatures, abundances, and area-scaled normalizations) were constrained to be equal across all datasets, with two exceptions.  First, the normalization of the source (e.g., filament emission) component was fixed at zero in background regions that did not include source emission.  Second, while the CXB normalizations were set equal for the on-axis Chandra regions, they were independent of the CXB normalizations for the I6 and {\it RASS} spectra, which were in turn independent of one another.  This was done to account for the different point source detection thresholds on axis, off axis, and with {\it ROSAT} (see Section~\ref{subsec:chsys}).

\subsection{Systematics Regarding the Chandra X-Ray Background} \label{subsec:chsys}

The ACIS-I6 chip, indicated by the 5 offset single CCDs from the primary observations in Figure~\ref{fig:mosaic}, is far from the telescope axis, while the four ACIS-I0-3 chips are relatively close to on the telescope axis. {\it Chandra} resolves point sources well, but still only to a limiting flux, which is different for on versus off axis observations. For the faint, diffuse emission that is characterized in this work, accurate modeling of the cosmic x-ray background (CXB) is essential. The fainter, unresolved point sources contribute flux that needs to be characterized. For the on-axis observations, we adopt the methodology for estimating the total flux from the unresolved CXB that \cite{bautz}, \cite{bulbul} and \cite{walker2012} implement in similar analyses using {\it Suzaku} data, which we describe here. 

The {\it Chandra} filament observations allow us to detect point sources down to a flux of $1.4_{-0.6}^{+2.6} \times 10^{-15}$ ergs cm$^{-2}$~s$^{-1}$, the faintest point source detected in our observations. \cite{moretti} defines the unresolved CXB flux in ergs cm$^{-2}$~s$^{-1}$~deg$^{-2}$ as:  

\begin{equation}
F_{CXB} = (2.18 \pm 0.13) \times 10^{-11} - \int_{S_{lim}}^{S_{max}} \frac{dN}{dS}\times S~ds.
\end{equation}
The analytic form of the source flux distribution in the 2-10 keV band is characterized as:

\begin{equation}
N(>S) = N_0 \frac{(2 \times 10^{-15})^\alpha}{S^\alpha + S_0^{\alpha - \beta}S^\beta} ~ \rm{ergs}~\rm{cm}^{-2}~\rm{s}^{-1}, 
\end{equation}
where $\alpha = 1.57_{-0.18}^{+0.10}$ and $\beta = 0.44^{+0.12}_{-0.13}$ are the power law indices for the bright and faint components of the distribution respectively, $N_0 = 5300_{-1400}^{+2850}$, $S_{lim}$ is the flux of the faintest point source detected in our filament observations, and $S_{max}$ is $8 \times 10^{-12}$ ergs cm$^{-2}$~s$^{-1}$. We find the unresolved CXB in our observations has a flux of $7.5 \pm 2.1 \times 10^{-12}$ ergs cm$^{-2}$~s$^{-1}$~deg$^{-2}$. 

Finally, the expected $1\sigma$ uncertainty in the unresolved CXB flux may be given by:

\begin{equation}
\sigma^2 = \frac{1}{\Omega}\int_{0}^{S_{lim}}\frac{dN}{dS}\times S^2~ds
\end{equation}
where $\Omega$ is the solid angle \citep{bautz}. We find the expected RMS deviation to be $1.5\times10^{-12}$ ergs cm$^{-2}$~s$^{-1}$~deg$^{-2}$.

We fix the on-axis CXB normalization to this unresolved flux and allow the normalization to vary within $1\sigma$. The off-axis CXB flux is well modeled without injecting such priors into the fits, so the off-axis CXB normalization is left free to vary independently in all {\it Chandra} fits.

\subsection{XMM-Newton Data Reduction and Analysis} \label{xmmda}

For the { \it XMM-Newton} data, we gather photon events registered by the MOS1, MOS2, and PN detectors of European Photon Imaging Camera (EPIC). To reduce any contamination of the photon detections by soft protons, solar flare periods are suppressed through a wavelet filtering of two event light curves extracted in the 10-12 and 1-5 keV energy band, respectively. 

\begin{figure}[ht!]
    \centering
    \includegraphics[width=0.45\textwidth]{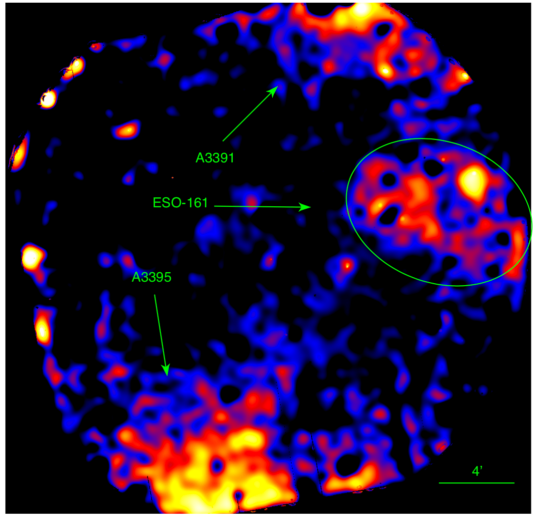}
    \caption{Log-scale {\it XMM-Newton} image of the A3391/A3395 intercluster filamentary region smoothed by a 19.6\arcsec~gaussian. This image is taken in the 0.3-2.5 keV band with MOS1, MOS2, and PN. The elliptical group region shown in Figure~\ref{fig:mosaic} (left panel) is shown for reference. Detected point sources are masked.}
    \label{fig:xmmpic}
\end{figure}
The exposure times after filtering are shown in Table~\ref{tab:pointings}. Events registered by anomalously bright CCDs of the MOS cameras \citep{kuntz2008} have also been suppressed. All events are rebinned spatially and spectrally into a cube that samples the mirror Point Spread Functions (PSF) and the detector energy responses. The resulting spatial binning is 1.6\arcsec, while the spectral binning increases in the range 15-190 eV as a function of event energies. Following the approach presented in \cite{bourdin2008}, effective exposure and background noise values are associated with each bin of the event cube and are subsequently used for both imaging and spectroscopy. The background noise model includes false detections of instrumental origin (detector fluorescence lines), cosmic induced particle background, and extended emission of astrophysical origin (CXB, and Galactic trans-absorption emission; see  \cite{kuntz2000}). More precisely, spatial and spectral variations of the instrumental background are modeled for each detector following the approach described in \cite{bourdin2013}. Amplitudes of the astrophysical emissions have been jointly fitted with the instrumental background in a sky region located to the northeast of the {\it XMM-Newton} pointing, which is spatially separated from the A3395-A3391 intercluster filament.

Spectroscopic measurements similarly rely on modeling the source emission measure provided by the APEC model. We similarly assume elemental abundances follow the solar composition tabulated in \cite{grevesse} and the Spectral Energy Distribution (SED) of the intercluster plasma is redshifted to z=0.0530. The ICM, GH, and CXB are absorbed assuming the Galactic Hydrogen column density reported in Section~\ref{subsec:chdra}. The LHB is  unabsorbed. For the CXB, the power-law index was fixed at 1.4, while the temperatures of the LHB and GH were fixed at 0.1 keV and 0.3 keV respectively. The normalizations are free to vary. The background model was fit simultaneously with the source model and the normalizations are generally consistent with the {\it Chandra} background normalizations within $2\sigma$ (see Table~\ref{tab:bg}). In this modeling, all astrophysical components are corrected for spatial variations of the instrument effective area and redistributed as a function of the energy response of the detectors.

Photon images and surface brightness profiles are corrected for the background noise model and the effective exposure time expected within their energy band. For these purposes, effective exposures assume the incidental photon energy to follow the SED of an isothermal plasma of temperature kT=5 keV. To increase the S/N, the exposure and background corrected photon image in Figure~\ref{fig:xmmpic} has been smoothed by a gaussian kernel of width $\sigma$(fwhm) = 19.2\arcsec. 

\begin{figure}[ht!]
\centering
\includegraphics[width=0.45\textwidth]{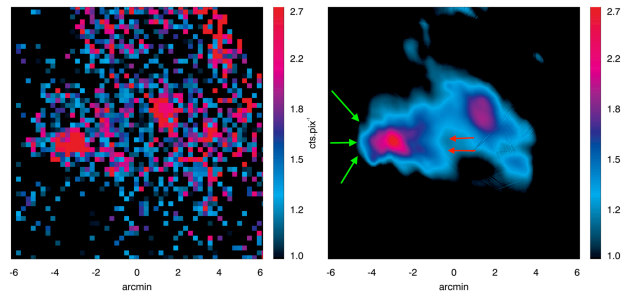}
\caption{{\it XMM-Newton} images of ESO-161 in the 0.3-2.5 keV energy range.  The color bar is counts/pixel. Left: The original {\it XMM-Newton} photon image of ESO-161 with 14.4\arcsec pixels. Right: Curvelet denoised at the $3\sigma$ level {\it XMM-Newton} image of ESO-161 derived from the left photon image. The green arrows indicate the eastern leading edge. The red arrows indicate the downstream edge discussed in Section~\ref{subsec:eso161}.}
\label{fig:coldfront}
\end{figure}

The image presented in Figure~\ref{fig:coldfront} uses curvelet denoising to preserve surface brightness edges. Specifically, a curvelet transfrom of first generation \citep{candes2002,starck} is computed from the photon image (see the left panel of Figure~\ref{fig:coldfront}). This transform combines ridgelet and wavelet bands, whose variance is stabilized following the Multiscale Variance Stabilized Transform proposed in \cite{Zhang}. Variance stabilized coefficients of the exposure corrected photon image are subsequently thresholded at 3$\sigma$, which yields a boolean support of significant coefficients. To restore the source surface brightness, a curvelet transform of the background noise image is projected onto the significant coefficient support and subtracted from the thresholded transform of the photon image shown right panel of Figure~\ref{fig:coldfront}.

\section{Results} \label{sec:results}

\subsection{Imaging} \label{subsec:imaging}

\begin{figure}[ht!]
    \centering
    \includegraphics[width=0.45\textwidth]{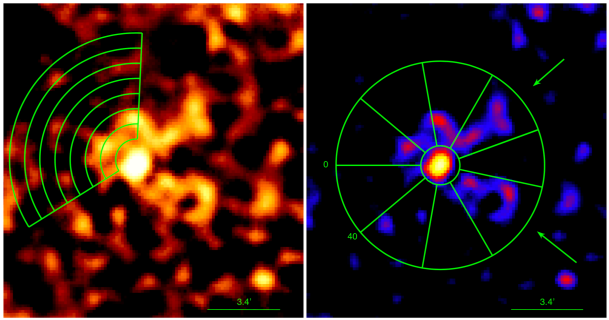}
    \caption{Close up \textit{Chandra} image of ESO-161 of Figure~\ref{fig:mosaic} (right panel). Left: ESO-161 with the annuli used for a surface brightness profile (see Section~\ref{subsec:eso161}) overlaid. Right: The overlaid azimuthal region is used for a surface brightness profile (see Section~\ref{subsec:eso161}). The arrows indicate where the two possible stripped gas tails are located.}
    \label{fig:eso161zoom}
\end{figure}

A3391 is the northern cluster of the system, with A3395 located to the
south. There is an extended filament of hot X-ray gas connecting these two clusters (see Figure~\ref{fig:xmmpic}). A3395 is comprised of two main subclusters, with a gas filament connecting the two subclusters in the east-west direction in the northern part of the system (see ObsID 4944 in the right panel of Figure~\ref{fig:mosaic}). ESO-161 is most clearly seen in Figures~\ref{fig:coldfront} and~\ref{fig:eso161zoom}, along with the extended diffuse
emission in Figure~\ref{fig:xmmpic} indicated inside the group ellipse region. 

The bridge of emission spanning $\sim 3$~Mpc in projection and connecting the subclusters is highlighted in Figure~\ref{fig:mosaic} (right panel) and Figure~\ref{fig:xmmpic} (for a wider field of view, see \cite{tittley,planck2013}). The box regions shown in Figure~\ref{fig:mosaic} (left panel) are used for the temperature, abundance, density, and entropy profiles derived in Sections~\ref{subsec:spectroscopy} and~\ref{subsec:entropyprofile}.

The galaxy group ESO-161 found in the intercluster filament has extended emission mostly to the west, which can be seen in Figures~\ref{fig:xmmpic}, and~\ref{fig:eso161zoom}. 

\begin{figure}[ht!]
    \centering
    \includegraphics[width=0.45\textwidth]{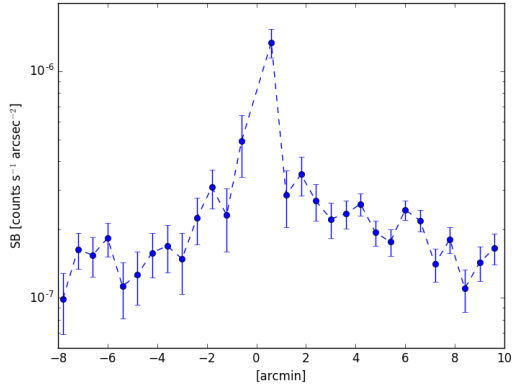}
    \caption{Surface brightness profile in the 0.3-2.5 keV band of ESO-161 extracted from the \textit{Chandra} observations with the wedges shown in the right panel of Figure~\ref{fig:mosaic}. Errors are $1\sigma$.}
    \label{fig:esosb}
\end{figure}

We created a surface brightness profile (Figure~\ref{fig:esosb}) in the east-west direction across ESO-161 from the region marked by wedges in Figure~\ref{fig:mosaic} (right panel). Here, the annuli are equally spaced bins of 0.6\arcmin. Note that the x-axis origin, 0\arcmin, of the surface brightness profile corresponds to the center of the annuli, with the west annuli noted as positive arcminutes and the east annuli are negative arcminutes. The goal of this analysis was to discern where the group emission reached the background emission. Therefore, a cut was made where the surface brightness profile flattens out at $\approx$ 8\arcmin~to the west and at $\approx$ 3\arcmin~to the east. The {\it XMM} image was then examined to refine the group region by eye as it is shown in Figure~\ref{fig:mosaic} (left panel). The group emission is elliptical and slightly angled in the NE-SW direction.

\subsection{Spectroscopy} \label{subsec:spectroscopy}

\begin{figure}[ht]
    \centering
    \includegraphics[width=0.45\textwidth]{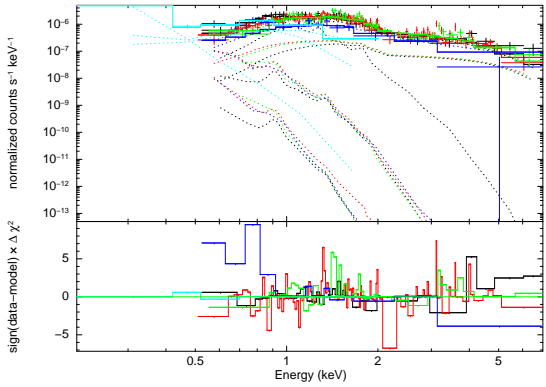}
    \caption{Top: The best fit spectrum for the box shown in the right panel of Figure~\ref{fig:mosaic}. The black, red, and green lines are from ObsIDs 13525, 13519, and 13522 respectively. The dark blue line is the simultaneously-fitted background spectrum for the dashed region on ObsID 13525 shown on the ACIS-I6 chip in the left panel of Figure~\ref{fig:mosaic}. The light blue line is the simultaneously-fitted background \textit{RASS} spectrum. Bottom: Residuals for the top spectra. Dotted lines are model components.}
    \label{fig:bigboxspec}
\end{figure}

To measure a gas mass, electron density, and temperature for the whole filament, we use the box region shown in the right panel of Figure~\ref{fig:mosaic} to extract the spectra in Figure~\ref{fig:bigboxspec}. The black, red, and green lines are the source spectra corresponding to ObsIDs 13525, 13519, and 13522 respectively. The dark blue line is the simultaneously fitted local background spectrum for the green dashed region on the ACIS-I6 chip for ObsID 13525 in the left panel of Figure~\ref{fig:mosaic}. The light blue line is the \textit{RASS} annulus simultaneously fitted background spectrum. We note the soft excess in the residuals for the local {\it Chandra} background component. Adding a soft proton component to the model does not improve the fit. Letting the GH and LHB parameters vary untied between spectra removes this soft excess for the I6 background spectrum in the residuals, however then the GH and LHB model parameters are then in tension with each other. Due to the low S/N of the I6 spectrum ($\sim 15\%$), we choose to leave these parameters tied. Performing either of these analyses, however, does not significantly change the best fit parameter values or error ranges.

\begin{figure}[ht!]
    \centering
    \includegraphics[width=0.45\textwidth]{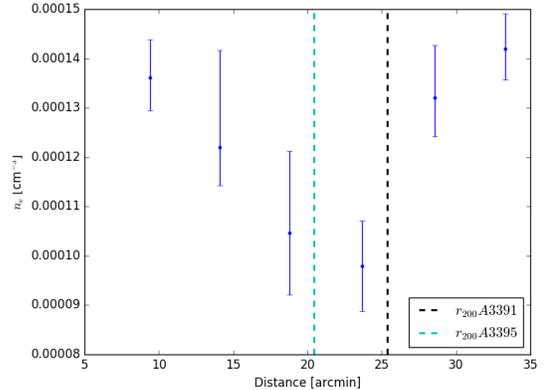}
    \caption{The electron density profile for the filamentary region derived from the 1T model normalizations. The black and cyan dashed lines are $r_{200}$ for A3391 and A3395 respectively. The X-axis is the distance from the center of A3391.}
    \label{fig:neprof}
\end{figure}

For all reported gas masses, we assume a 3D cylindrical geometry for the filament with the length and radius dimensions corresponding to the box region length and half-width edges respectively, assuming that the filament is slightly more extended than what is encompassed within the 16\arcmin~by 16\arcmin~{\it Chandra} field of view. The box region in Figure~\ref{fig:mosaic} (right panel), for a 3D cylindrical geometry has a radius of 0.7 Mpc and a length of 0.9 $\sin(i)^{-1}$~Mpc, where $i$ is the inclination angle of the filament to the line of sight. The box region was placed where the filament emission is relatively bright and where the ICM emission is relatively faint, thus maximizing the S/N of the filament emission. The box regions used for the profiles in Figure~\ref{fig:mosaic} (left panel) were assumed to have a 3D cylindrical geometry with radii of 0.7 Mpc and a length of 0.3 $\sin(i)^{-1}$~Mpc. The electron density of the filament is derived from the normalization in XSPEC, and is given by:
\begin{equation}
\begin{split}
    n_e  = \biggl(1.6 \ N \ \rm{sin}({\it i}) \times 1.08\times10^{-10}(1+z)^2 \\ \biggl (\frac{D_A}{\rm{Mpc}}\biggl)^2\biggl(\frac{r}{\rm{Mpc}}\biggl)^{-2}\biggl(\frac{l_{obs}}{\rm{Mpc}}\biggl)^{-1}\biggl)^{1/2},
\end{split}
\label{eq:electrondensity}
\end{equation}
where $D_A$ is the angular distance to the system, $r$ is the radius
of the filament, $l_{obs}$ is the observed 
projected
length of the filament, and $N$ is the XSPEC normalization. The electron density profile across the filament is shown in Figure~\ref{fig:neprof}.

The aforementioned box region (see right panel of Figure~\ref{fig:mosaic}) was best fit with a two temperature APEC model ($\chi^2/dof = 619.46/594$) rather than a one temperature APEC model ($\chi^2/dof = 714.96/595$). We find projected temperatures kT = $4.45_{-0.55}^{+0.89}$~keV and $0.29_{-0.03}^{+0.08}$~keV. Under the assumption that the hotter temperature component is that associated with the filament (see Section~\ref{subsec:filament} for discussion), we find an electron density $n_e=1.08^{+0.06}_{-0.05} \times 10^{-4}$~$\sin(i)^{\frac{1}{2}}$~cm$^{-3}$, and $M_{\rm gas} = 2.7^{+0.2}_{-0.1} \times 10^{13}$~$\sin(i)^{-\frac{1}{2}}$~M$_\odot$ for the filament assuming that this inferred density extends outside the {\it Chandra} FOV into the box region shown in Figure~\ref{fig:mosaic}. If the filament is indeed completely covered by the {\it Chandra} FOV and is not extended, then the gas mass would be $\sim 1.7 \times 10^{13}$~$\sin(i)^{-\frac{1}{2}}$~M$_\odot$. This gas mass is in good agreement with \cite{tittley}. Note that our errors are statistical; there are additional systematic errors associated with the assumed cylindrical geometry and unknown substructure of the filament.

The mean baryonic density of the universe at $\bar{z} = 0.053$ is $\bar{\rho}_{baryon} \simeq 4 \times 10^{-31}$ g cm$^{-3}$. If the filament is in the plane of the sky (see Section~\ref{subsubsec:orientation}), $\frac{\rho_{fil}}{\bar{\rho}_{baryon}} \lesssim 541$. If the filament is indeed aligned close to the plane of the sky, this overdensity is not consistent with that expected for the WHIM gas, which is thought to range between 1-250 \citep{bregman}. 

\begin{figure}[ht!]
    \centering
    \includegraphics[width=0.45\textwidth]{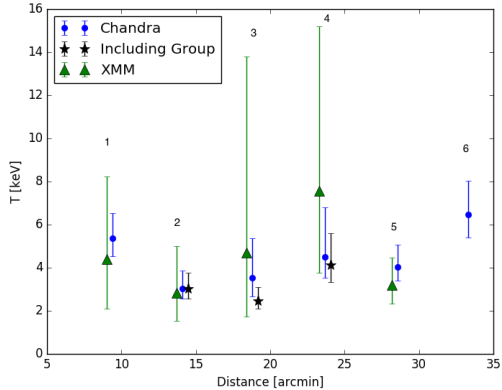}
    \caption{The projected temperature from \textit{Chandra} and \textit{XMM-Newton}
      observations as a function of distance from the center of the
      northern subcluster, A3391. Each point corresponds to a box
      region from which we extracted spectra (see the left panel of Figure~\ref{fig:mosaic}). The black points include the group emission and offset for
      viewing purposes. The green triangles are the \textit{XMM} measurements and the blue circles are the \textit{Chandra} measurements. Regions are labeled for reference.}
    \label{fig:tempprof}
\end{figure}

The boxes shown in Figure~\ref{fig:mosaic} (left panel) are used to extract spectra and
create the projected one temperature (1T) profile. In Figure~\ref{fig:tempprof}, we display
three temperature profiles; two {\it Chandra} profiles including the group and excluding
the group ESO-161, and the {\it XMM} profile also excluding the group region. 

We additionally fit a two temperature (2T) APEC model for the same spectra in each of the regions in the profile, keeping the metallicity fixed at $0.3 Z_\odot$. We find that fitting a second cool component to the spectrum of the northern filament observation only for regions 2 and 3 yields better statistically significant 2T fits shown in Table~\ref{tab:2temp} (see Section~\ref{subsec:filament} for discussion).

\begin{table}[]
\centering
\caption{The best fit parameters for 1 and 2 temperature models for regions 2 and 3 in the 0.5-7.0 keV band.}
\label{tab:2temp}
\begin{tabular}{@{}lccccc@{}}
\toprule
Reg & \begin{tabular}[c]{@{}c@{}}kT$_1$\\ {[}keV{]}\end{tabular} & \begin{tabular}[c]{@{}c@{}}$N_1$\\ ($10^{-4}$)\\ {[}cm$^{-5}${]}\end{tabular} & \begin{tabular}[c]{@{}c@{}}kT$_2$\\ {[}keV{]}\end{tabular} & \begin{tabular}[c]{@{}c@{}}$N_2$\\ ($10^{-4}$)\\ {[}cm$^{-5}${]}\end{tabular} & $\chi^2$/dof \\ \midrule
2      & $3.03_{-0.49}^{+0.88}$                                     & $9.38_{-0.50}^{+0.10}$                                                        & ...                                                        & ...                                                                           & 359.51/328   \\
2      & $4.77_{-1.21}^{+2.24}$                                     & $7.27_{-0.71}^{+0.68}$                                                        & $0.44_{-0.10}^{+0.17}$                                     & $6.74_{-2.58}^{+3.23}$                                                        & 301.83/327   \\
3      & $3.53_{-0.87}^{+1.85}$                                     & $7.10_{-0.64}^{+0.84}$                                                        & ...                                                        & ...                                                                           & 203.55/211   \\
3      & $3.87_{-1.00}^{+2.46}$                                     & $7.15_{-0.69}^{+1.24}$                                                        & $0.28_{-0.09}^{+0.15}$                                     & $3.89_{-2.47}^{+8.49}$                                                        & 183.87/210   \\ \bottomrule
\end{tabular}
\end{table}

\begin{figure}[ht!]
    \centering
    \includegraphics[width=0.45\textwidth]{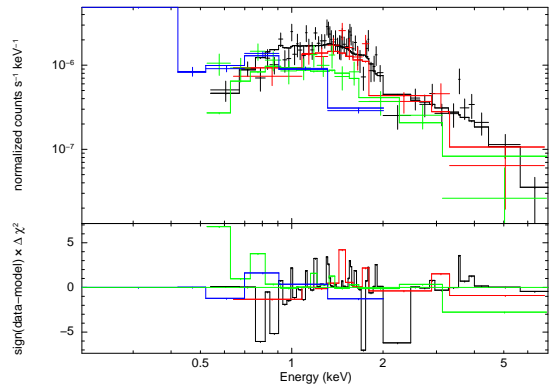}
    \caption{Spectral fit and the residuals for Region 4, located approximately at $r_{200}$ for both clusters, shown in the left panel of Figure~\ref{fig:mosaic}. The black and red lines are the source spectra from north to south respectively, the green line is the simultaneously fitted dashed background region (see the left panel of Figure~\ref{fig:mosaic}), and the blue line is the simultaneously fitted background \textit{RASS} spectrum.}
    \label{fig:allfilm2spec}
\end{figure}

We find a best fit 1T model with projected temperature kT $= 4.49^{+2.31}_{-0.97}$ keV, and electron density $n_e = 9.79^{+0.92}_{-0.92} \times 10^{-5}$~$\sin(i)^{\frac{1}{2}}$~cm$^{-3}$
for Region 4. This is the only area in our study that lies approximately at $r_{200}$ of both of the subclusters. Figure~\ref{fig:allfilm2spec} shows the spectrum for Region 4, where the black (13519) and red (13522) lines are the source spectra, the green line (13525) is the simultaneously fit dashed background region (see the left panel of Figure~\ref{fig:mosaic}), and the blue line is the simultaneously fit background \textit{RASS} spectrum.

We obtained upper limits of the projected abundances (Table~\ref{tab:abund}) for regions 1, 3, 4, and 6 shown in the left panel of Figure~\ref{fig:mosaic} derived from an absorbed APEC model.

\begin{table}[H]
\centering
\caption{Projected abundance values for the box regions shown in Figure~\ref{fig:mosaic} (left panel).}
\label{tab:abund}
\begin{tabular}{ll}
\hline
Region & \begin{tabular}[c]{@{}l@{}}Abundance {[}$Z_\odot${]}\end{tabular} \\ \hline
1                    & $<0.72$                                                                      \\
2                    & $0.14^{+0.12}_{-0.12}$                                                                       \\
3                    & $<0.63$                                                                      \\
4                    & $<0.65$                                                                                    \\
5                    & $0.60^{+0.74}_{-0.20}$                  \\
6                    & $<0.63$ \\ \hline
\end{tabular}
\end{table}

The temperature profile along the intercluster filament derived from {\it XMM} observations is in good agreement within uncertainties with the Chandra results. We could not fit Region 6 with the {\it XMM} data, as the region is only covered by EMOS1 after the filtering of bright CCDs, and the signal to noise is too low to constrain the fit. The large uncertainties for Region 3, and to a lesser extent Region 4, are a result of excluding the group region and subsequently having low areal coverage of the observation (see the left panel of Figure~\ref{fig:mosaic}), as well as a low inherent S/N as Region 4 is our faintest region. Our temperature, abundance, and density profiles (see Section~\ref{subsec:entropyprofile}) are also in good agreement with those found with {\it Suzaku} in \cite{sugawara}.

\section{Discussion} \label{sec:discussion}

\subsection{The Filament}\label{subsec:filament}

Here we discuss the derived density and entropy profiles, as well as the galaxy group ESO-161 to further explore the orientation and nature of the system.

\subsubsection{Nature of the Filament} \label{subsec:entropyprofile}

We fit the filament data with 2T models because there is likely contamination from ICM emission between $r_{500}$-$r_{200}$. As shown in Section~\ref{subsec:spectroscopy}, we find that a 2T model is a better statistically significant fit for regions 2 and 3 with a cooler component ranging from $\sim 0.2-0.6$~keV (see Table~\ref{tab:2temp}). We find a group temperature of $\sim 1.09$ keV (see Section~\ref{subsec:eso161}). We find that the measured cooler components in the filament regions are consistent with the temperature profile one would expect for an $\sim 1$~keV group at this distance from the group center based on the universal group temperature profile derived by \cite{sun}. Therefore, the 2T fits for both of the aforementioned regions as well as the box region shown in the right panel of Figure~\ref{fig:mosaic}, which all cover $r_{500}$ for the group, indicate that there is extended group emission in the filament beyond the group excluded region shown in Figure~\ref{fig:mosaic} (left panel).

The electron density (see Equation~\ref{eq:electrondensity}) profile for the filament, assuming it is in the plane of the sky, is shown in Figure~\ref{fig:neprof}. The black and cyan dashed lines are $r_{200}$ for A3391 and A3395 respectively. There appears to be a dip in the electron density at the midpoint of the filament, at $\sim r_{200}$ of both the subclusters. This minimum in the density profile is approximately 2 dex higher than the mean baryonic density of the universe at the mean redshift of the system.

\begin{figure}[ht!]
    \centering
    \includegraphics[width=0.5\textwidth]{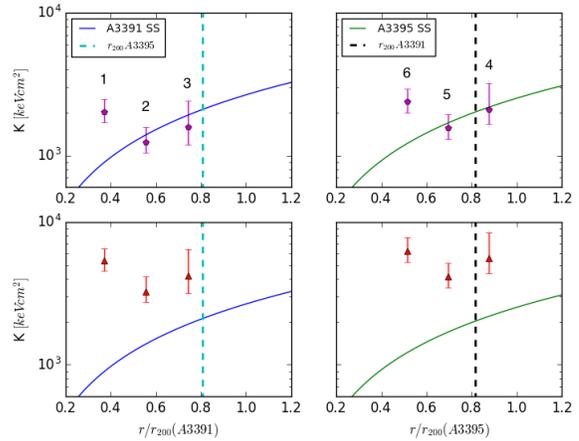}
    \caption{The blue line shows the self-similar entropy profile for A3391 derived from \cite{voit} (left two panels). The green line is the same, but for A3395 (right two panels). The black vertical dashed line is $r_{200}$ for A3391, and the cyan vertical dashed line is $r_{200}$ for A3395. The data points are the derived entropy for the green box regions shown in the left panel of Figure~\ref{fig:mosaic} and points are labeled for reference. The magenta pentagons are the entropy, assuming the filament is in the plane of the sky, and the red triangles are entropy values for a filament orientation i = $3.1^\circ$ to the line of sight, as suggested by \cite{tittley}. The distance shown on the x-axis is the distance from the cluster center.}
    \label{fig:entropyprofile}
\end{figure}

The entropy profile is shown in Figure~\ref{fig:entropyprofile} where we define entropy as $K = k_BTn_e^{-2/3}$, where $k_b$ is Boltzmann's constant, $n_e$ is the electron density, and $T$ is the temperature. The entropy profiles for galaxy clusters derived from \cite{voit} for A3391 and A3395 are shown in Figure~\ref{fig:entropyprofile}, where the center of each cluster was determined from NED. The blue and green lines are the self-similar entropy profiles for A3391 and A3395 respectively:
\begin{equation}
K(r) = 1.41 \pm 0.03~K_{200}~(r/r_{200})^{1.1}.
\end{equation}
$K_{200}$ is the entropy at $r_{200}$ and is defined as:
\begin{equation}
\begin{split}
K_{200} = 362~\rm{keV}~\rm{cm}^2~\frac{T_X}{1~ keV}~\biggl(\frac{T_{200}}{T_X}\biggl) \\
\times\biggl(\frac{H(z)}{H_0}\biggl)^{-4/3}\biggl(\frac{\Omega_m}{0.3}\biggl)^{-4/3},
\end{split}
\end{equation}
where $T_X \approx T_{200}$, and $\Omega_m$ is the matter density parameter. The black vertical dashed line is $r_{200}$ for A3391, and the cyan vertical dashed line is $r_{200}$ for A3395. $r_{200}$ values for the clusters were estimated using reported values of $r_{500}$ \citep{piffaretti} and assuming $r_{200} \sim 1.7r_{500}$ \citep[e.g.][]{ami}. We note that these values for $r_{200}$ are slightly smaller than those reported in \cite{sugawara}. The values for $r_{200}$ derived in \cite{sugawara} are estimated using the empirical $r_{200}-T_X$ relation \citep{henry} and are 2.3 and 2.1 times the measured $r_{500}$ value for A3391 and A3395 respectively. This difference in $r_{200}$ affects the normalization of the self-similar entropy profile (see Section~\ref{subsubsec:orientation} for discussion).

The data points in Figure~\ref{fig:entropyprofile} represent the entropy derived from the measured gas temperatures and electron densities shown in Figure~\ref{fig:tempprof} and Figure~\ref{fig:neprof} respectively. The 90\% temperature and electron density errors were propagated to derive the 90\% entropy uncertainties (see Figure~\ref{fig:entropyprofile}). The magenta pentagons are the entropy of the filament assuming it is in the plane of the sky ($i = 90^\circ$) and the red triangles are entropy values for a filament orientation $i = 3.1^\circ$ to the line of sight, the lowest inclination to the line of sight that \cite{tittley} argue for following their dynamical analysis of the system.

Even assuming the filament is in the plane of the sky, the entropy at large radii, namely near $r_{200}$ for both clusters, is much larger than the expected entropy values for the dense range of the WHIM gas by at least a factor of four, with the predicted value for the WHIM at this redshift being approximately 250 keV cm$^2$ \citep{vala}.

All of the regions for the profile lie inside $r_{200}$ of one, or both for the case of Region 4, of the subclusters. The extended ICM gas is expected to be hotter than the WHIM, and will bias the electron density towards higher values, so it is unclear what the overall entropy bias is due to these regions overlapping with the subcluster outskirts.

The radius of the filament profile geometry was assumed based upon the size of the \textit{Chandra} observations, and the filament may actually be more extended than what is captured in the 16\arcmin~by 16\arcmin~observations. If this is the case, our electron density measurements are biased high. This in turn biases the entropy low, reinforcing the conclusion that the gas in the filamentary region is ICM gas, as the entropy across the filament is already too high to be consistent with the WHIM emission.

We find no evidence for a shock that would support the suggestion by \cite{sugawara} of shock heated gas in this region. The flat temperature profile across the filament is consistent with ICM gas undergoing tidal pulling into the filament due to an early stage merger between the clusters. An interaction between the subclusters was also recently suggested by \cite{sugawara}.

\subsubsection{Orientation of the Filament}\label{subsubsec:orientation}

It has been shown that the entropy profile of most massive clusters lies at about self-similar within $r_{500}$, and then flattens below self-similar at larger radii \citep[i.e.][]{walker2013}. There is an uncertainty when relating a measured $r_{500}$ with $r_{200}$. The effect that this has on the self-similar entropy profile is in the normalization of the profile. This uncertainty in the normalization of the self-similar entropy profile makes it difficult to say with conviction which inclination brings the profile closer to an expected self-similar value. Figure~\ref{fig:entropyprofile} suggests a filamentary geometry close to the plane of the sky. The larger $r_{200}$ values determined empirically by \cite{sugawara} only serve to strengthen the argument for a filament orientation closer to the plane of the sky than the range close to the plane of the sky reported in other works, as the larger $r_{200}$ values decrease the normalization of the self-similar entropy profile. In any case, this uncertainty in normalization does not change the observed flattening of the entropy profile.

Given that the global filamentary temperature is $\sim 4.5$ keV,  the gas is very likely from the ICM outskirts of the two clusters, in which case the clusters must be close enough to be tidally interacting and cannot have a large line-of-sight separation.

\cite{tittley} 
found through a
dynamical analysis of the system that the subclusters and the connecting filament are oriented close to
the line of sight, having an inclination, $i$, between $3.1^\circ-9.0^\circ$. \cite{sugawara} suggest that the filament may be inclined $\approx 10^\circ$ to the line of sight in order for their X-ray measured Compton $y$ parameter to agree with the $y_{SZ}$ parameter reported by \cite{planck2013} for the filament. However, \cite{sugawara} also suggest that the discrepancy in $y$ parameters is likely a combination of the system not being in the plane of the sky, or there is unresolved multi-phase gas or shock heated gas present in the {\it Suzaku} observations. Indeed, if the system is inclined $10^\circ$ to the line of sight, the true separation between the subclusters would be over 17 Mpc, making it unlikely for the clusters to be interacting. However, we note that the center of the galaxy group ESO-161 is just outside the {\it Suzaku} field of view, so the extended cooler phase gas from the group, mixing with the surrounding filament gas may be an explanation for the $y$ parameter discrepancy.

The line of sight velocity difference of the clusters ($\sim 240$ km s$^{-1}$ \citep{struble}) is rather small and consistent with an early stage merger without a large line-of-sight peculiar velocity component. If the velocity difference were much larger then that would imply the clusters are significantly unbound and unable to interact tidally, or that the clusters are undergoing a late stage merger. The former scenario is in contradiction with the temperature and entropy values that we measure, and the latter scenario contradicts the observed line of sight separation between the clusters.

\subsection{ESO-161} \label{subsec:eso161}

To constrain the temperature of the galaxy group ESO-161, we
 fit the group region (see the left panel of Figure~\ref{fig:mosaic}) with an
 absorbed APEC model and the same background prescription described in
 Section~\ref{sec:da}; we use the dashed circular region to the east of the group shown in the left panel of Figure~\ref{fig:mosaic} to simultaneously model the local background.
 Our fit yielded a temperature of $1.09^{+0.58}_{-0.05}
 \rm{keV}$ for the group. This temperature is significantly cooler than the best fit for the surrounding region, $4.45^{+0.89}_{-0.55}$ keV (see Section~\ref{subsec:spectroscopy}). 

\begin{figure}[ht!]
    \centering
    \includegraphics[width=0.45\textwidth]{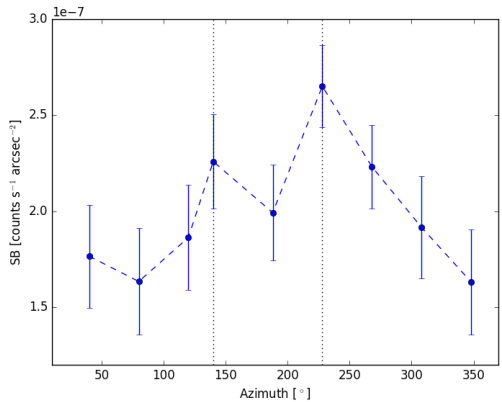}
    \caption{The surface brightness profile for the azimuthal regions shown in Figure~\ref{fig:eso161zoom} in the 0.3-3.0 keV band. The dotted lines indicate the position of the arrows in Figure~\ref{fig:eso161zoom}, pointing towards the visual stripped tails of gas. Errors are $1\sigma$.}
    \label{fig:esoazprof}
\end{figure}

The emission to the west of the group (see Figure~\ref{fig:xmmpic}) is indicative of a diffuse tail. With \textit{Chandra}, this diffuse gas can be resolved into a bimodal structure (see Figure~\ref{fig:eso161zoom}). We use the azimuthal regions shown in Figure~\ref{fig:eso161zoom} to derive the azimuthal surface brightness profile in the 0.3-3.0 keV band shown in Figure~\ref{fig:esoazprof}. This profile shows a hint of a double peak, in the same position as the arrows pointing towards the ram pressure stripped tail candidates (see the right panel of Figure~\ref{fig:eso161zoom}) indicated by the dotted lines, furthering evidence that the tail indeed may have a bimodal structure. The extended emission to the west of ESO-161 is suggestive that the
group may be undergoing ram pressure stripping as the group moves
through the filament. 

The bimodal tail may have a ``downstream edge" to the west of the group center, which is more apparent in the right panel of Figure~\ref{fig:coldfront}. This bimodal tail structure may indicate an ellipsoidal potential in origin \citep[e.g.][]{roediger} for the group. \cite{randallm86} first suggested that the double tails are due to stripping  from ellipsoidal potentials.

Another clue bolstering the ram pressure stripping scenario is the possible cold front shown in Figure~\ref{fig:coldfront}. The northeastern edge is consistent with the ``upstream edge" reported in \cite{roediger} for systems experiencing ram pressure stripping as they move through an ambient medium.

\begin{figure}
    \centering
    \includegraphics[width=0.45\textwidth]{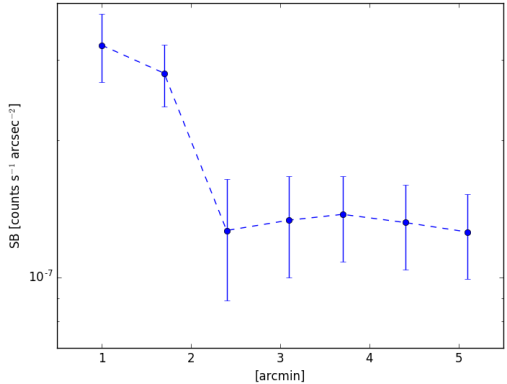}
    \caption{The {\it Chandra} surface brightness profile in the 0.3-3.0 keV band for the region shown in the left panel of Figure~\ref{fig:eso161zoom}, centered on the group. Errors are $1\sigma$.}
    \label{fig:coldfrontSB}
\end{figure}

To further investigate the prominent edge seen to the east of the galaxy group in Figure~\ref{fig:coldfront}, we derived a surface brightness profile in the northeast region of the group shown in Figure~\ref{fig:eso161zoom} (left panel), which may be seen in Figure~\ref{fig:coldfrontSB}.

There is a clear drop in surface brightness at $\sim$2\arcmin~ in Figure~\ref{fig:coldfrontSB}, moving radially away from the group. We do not have enough data to distinguish if this edge is really a cold front, shock front, or neither. More observing time with {\it XMM-Newton} would shed light on this question. This apparent edge, as well as the bimodal tail structure of stripped gas is indicative that the galaxy group is consistent with moving east in projection through the filament.

\cite{gg} give the conditions for ram pressure stripping to occur as
$P_{ram} = \frac{1}{2}\rho_{ICM}v_r^2~\gtrsim~\sigma^2 \rho_{gas}$,
where $\rho_{ICM}$ is the density of the intercluster medium, $v_r$ is
the velocity of the group relative to the intercluster medium,
$P_{ram}$ is the ram pressure, $\sigma$ is the galaxy group's velocity
dispersion, and $\rho_{gas}$ is the galaxy's gas density. 
In order to estimate the gas density of the group, we assume an oblate spheroid geometry, with the line-of-sight axis equal to the projected major axis and the minor axis in the plane of the sky. Using the electron density inferred from the box region in Figure~\ref{fig:mosaic} (right panel), the density for the group region is
$n_e \sim 1.8 \times 10^{-4}$~cm$^{-3}$.
\cite{tittley} report that the group velocity dispersion is $1800$~km/s. This velocity dispersion is much too high for the group to be bound, so we use the following method to roughly estimate the velocity dispersion of ESO-161.

We use the $M_x-T_x$ relation derived by \cite{vikhlinin} to estimate $M_{500}$. While the sample used to derive the $M_x-T_x$ relation in \cite{vikhlinin} consists of galaxy clusters and not groups, \cite{sun} report that the relation also holds for lower temperature galaxy clusters and groups. We find that $M_{500} \sim 2.3 \times 10^{13}$~M$_\odot$. Assuming spherical symmetry we may then use
$M_{500} = 500 \times \rho_c \frac{4}{3} \pi r_{500}^3$,
where $\rho_c$ is the critical density of the Universe at the redshift of ESO-161, $9.86\times10^{-30}$~g cm$^{-3}$, and $r_{500}$ is the radius at which the density of the galaxy group is 500 times the critical density of the Universe at the redshift of the galaxy group, to estimate the radius of the group. Finally, we may then use
$M_{500} = \frac{3}{G}\sigma^2r_{500}$,
where $G$ is the gravitational constant and $\sigma$ is the velocity dispersion of the group. We find that the group has a velocity dispersion of $\sim 250$~km s$^{-1}$. This velocity dispersion is approximately six times lower than what \cite{tittley} report.

We find that the group must have a relative velocity to the filamentary region $v_r \geq 360$~km s$^{-1}$ in order for ram pressure stripping to occur. If the filament is oriented to the median inclination angle given by \cite{tittley}, then the minimum relative velocity would have to be $\sim 630$ km/s as the group moves eastward through the filament.

Another possibility for the extended emission to the west of the group is tidal stripping, perhaps due to a gravitational interaction with another massive object. The extended emission to the north of the group (see the right panel of Figure~\ref{fig:coldfront}), as well as the $<$1 keV temperatures found in regions to the north of the group (see Table~\ref{tab:2temp}) could indicate that ESO-161 is moving to the southeast, around A3391, and the group experienced a tidal stripping event. The emission to the west may also be the result of tidal stripping due to an interaction with a currently unidentified, possibly gas stripped object.

In any case, these gaseous double tail-like structures are commonly seen in ram pressure stripped galaxies, most notably in the Virgo Cluster \citep[i.e.][]{forman1979,randallm86}, and in simulations \citep[i.e.][]{roediger}. This would therefore lead to the conclusion that ESO-161 is being ram pressure stripped as it moves eastward in projection through the intercluster filament. We note that such clear examples of ram pressure stripped galaxy groups near low density cluster outskirt environments are quite rare \citep[e.g.][]{degrandi}.

\section{Summary and Conclusion} \label{sec:summary}

We have presented results based on {\it Chandra} and {\it XMM-Newton} observations of the intercluster gas filament connecting A3391 and A3395.  We find the following:

\begin{itemize}
\item[--] A global projected temperature kT = $4.45_{-0.55}^{+0.89}$~keV, electron density $n_e=1.08^{+0.06}_{-0.05} \times 10^{-4}\sin(i)^{\frac{1}{2}}$~cm$^{-3}$ for the intercluster filament.
\item[--] The temperature and electron density derived for the global intercluster filamentary region indicates that the filament gas mass is $M_{\rm gas} = 2.7^{+0.2}_{-0.1} \times 10^{13}\sin(i)^{-\frac{1}{2}}$~M$_\odot$. This is a similar mass to what is reported for the intercluster filament between the two subclusters Abell 222 and Abell 223 \citep{werner}, and is consistent with what \cite{tittley} found in their analysis with \textit{ROSAT} for A3395/A3391.
\item[--] The temperature and entropy profiles derived for the filament suggest ICM gas is being tidally pulled into the intercluster filamentary region as part of an early stage pre-merger. The density across the intercluster filament is consistent with the dense WHIM as well as what is expected in cluster outskirts, near the virial radius, although the temperature and entropy are much higher than what is expected for the WHIM.
\item[--] The galaxy group ESO-161, located between A3391 and A3395 in the intercluster filament, may be undergoing a stripping event as the group moves eastward seemingly perpendicular to the filament with a minimum relative velocity of approximately 360 km s$^{-1}$ if the filament is oriented in the plane of the sky. In addition, the group has a distinct edge in surface brightness to the east, which would require a deeper observation with {\it XMM-Newton} to characterize.
\end{itemize}

Since the subclusters appear to be tidally interacting, their line of sight separation must not be large, leading us to conclude that the filament is probably not oriented close to the line of sight as was suggested by \cite{tittley}. Furthermore, an in-plane orientation yields a density that is more consistent with ram pressure stripping of the galaxy group ESO-161.

The only evidence we find for cooler phase gas is that likely associated with the galaxy group ESO-161. The filament density, even in projection, is consistent with the theoretical density of the WHIM, however this density is also consistent with density profiles of clusters out to the virial radius \citep{morandi}.

\cite{sugawara} argue that the filament temperature is too high to be explained by universal cluster temperature profiles of the subclusters, and attribute this to a shock, perhaps as the subclusters merge. We do not find evidence for merger shocks in the filament. The 4.5 keV filament temperature that we measure is consistent with ICM gas being tidally pulled into the intercluster filament in the early stages of a massive cluster merger. This heated gas above the cluster temperature profiles as well as temperatures expected for the WHIM could also be attributed to adiabatic compression in the filament.

\acknowledgments

We would like to thank Rodolfo Montez Jr., and Yuanyuan Su for their enlightening discussions and comments. We would also like to thank the referee for their helpful comments. SWR was partially supported by the Chandra X-ray Center through NASA contract NAS8-03060, and the Smithsonian Institution.  Support for GEA was partially provided by Chandra X-ray Observatory grants GO2-13161X and GO3-14133X, and by the Graduate Assistance in Areas of National Need (GAANN). This publication received federal support from the Latino Initiatives
Pool, administered by the Smithsonian Latino Center. 
{}

\bibliographystyle{aasjournal}
\bibliography{33913395bib}

\end{document}